\documentclass[%
reprint,
amsmath, amssymb, 
longbibliography,
aps,
]{revtex4-1}

\usepackage{graphicx}
\usepackage{subfigure}
\usepackage{dcolumn}
\usepackage{bm}

\begin{document}

\preprint{APS/123-QED}
\title{Measuring entanglement of photons produced by a pulsed source}

\author{M\'onica B. Ag\"uero}
 \altaffiliation[Corresponding author: ]{maguero@citedef.gob.ar, \\ moguby@gmail.com}


\author{ Alejandro A. Hnilo}
\author{Marcelo G. Kovalsky}
\affiliation{CEILAP, Centro de Investigaciones en L\'aseres y Aplicaciones, UNIDEF (MINDEF-CONICET);\\
 J.B. de La Salle 4397, (1603) Villa Martelli, Argentina. 
}%


\date{\today}

\begin{abstract}
A pulsed source of entangled photons is desirable for some applications. Yet, such a source has intrinsic problems arising from the simultaneous arrival of the signal and noise photons to the detectors. These problems are analyzed and practical methods to calculate the number of accidental (or spurious) coincidences are described in detail, and experimentally checked, for the different regimes of interest. The results are useful not only to measure entanglement, but to all the situations where extracting the number of valid coincidences from noisy data is required. As an example of the use of those methods, we present the time-resolved measurement of the Concurrence of the field produced by spontaneous parametric down conversion with pump pulses of duration in the ns-range at a repetition of kHz. The predicted discontinuous evolution of the entanglement at the edges of the pump pulse is observed.
\begin{description}
\item[PACS numbers]
42.50.Xa, 42.65.Lm, 42.50.Dv. 
\end{description}
\end{abstract}

\pacs{Valid PACS appear here}
\maketitle

 \section{INTRODUCTION}
Entangled states of photons are the basic resource in the successful implementation of quantum information processing applications, namely optical quantum computing \cite{knill2001,walther2005}, and quantum cryptography, or quantum key distribution \cite{ekert1991,jennewein2000}. 
Also, in the experimental study of fundamental problems in Quantum Mechanics, as loophole-free tests of the violation of the Bell's inequalities \cite{giustina2013}, the delayed-choice quantum eraser \cite{kim2000}, quantum teleportation and entanglement swapping \cite{jennewein2001}, generation of states with a large number of particles and generalized types of entanglement \cite{zhang2006}, etc. 
The standard method for generating entangled photon states nowadays is spontaneous parametric down conversion (SPDC), which is achieved by pumping one or more nonlinear crystals with a laser source. The most common case corresponds to a continuous-wave (CW) pumping laser with a narrow bandwidth. Nevertheless, there are regimes of interest other than the CW one. Quantum computing is implicitly thought to operate with a ``clock'' to synchronize the different transformations to be applied to the quantum system. In quantum key distribution at large distances or in daylight, it is convenient knowing the possible time of arrival of the signal photons to reduce undesirable background light through temporal discrimination, and also as a defense against a sophisticated eavesdropper \cite{gerhardt2011}. In the area of the experimental study of the foundations of Quantum Mechanics, the knowledge of that time is necessary to close the so-called coincidence-loophole \cite{LG2004} and to test the hypotheses regarding the time of formation of the entanglement \cite{michielsen2011,hnilo2012}. In all these cases, a pulsed source of entangled photon pairs (or ``biphotons'', as D. Klyshko named them) is desirable and, in some cases, unavoidable. 

Pumping with femtosecond [fs] laser pulses is customary in the setups of entanglement swapping or teleportation, as well as to achieve an event-ready, or heralded, source of biphotons. It is convenient stressing here that an event-ready source (i.e., a heralded source of a Fock state of biphotons, which is a highly non-classical state of the field) cannot be achieved by merely pulsing the pump, but by combining several processes of two-photon SPDC \cite{sliwa2003}, or by one process of three-photon SPDC \cite{hnilo2005}. The achievable rate of heralded biphotons is in consequence very low. Fortunately, an event-ready source is not required for many of the applications or studies mentioned before. In these cases, it is not really necessary knowing the time when a biphoton is to be observed, but only the time when it \textit{cannot} be observed. For this purpose, a simple pulsed-pump source of low intensity (which produces a coherent, almost-classical state of the field) suffices. 

In principle, the pulsed-pump operation implies a finite (i.e., large) bandwidth. The problems due to the broad spectral bandwidth of the fs pulses have been analyzed in \cite{joobeur1994,grice1997,kim2002}. The case of picosecond [ps] pulse pumping, where the bandwidth is much narrower, has been considered in \cite{kuzucu2008}. Yet, in both cases the source is a mode-locked laser, where the pulse repetition rate is in the order of 100 MHz. The successive pulses are so close, that the time values of the valid photon detections are in the limit of what can be reliably processed (i.e.: detected, identified and saved) with the currently available devices. 
Well separated pulses are hence necessary to record the time values of photon detection. Besides, well separated pulses are needed to test the time-coincidence loophole \cite{LG2004}, and relatively long pulses allow the observation of the time variation of entanglement inside the pulse, which is an interesting and almost unexplored issue. On the other hand, the regime of well separated pulses has the evident practical problem that the signal (entangled) photons and the noise (uncorrelated) photons arrive to the detectors simultaneously. This implies that a large number of accidental or spurious coincidences occur simultaneous with the valid ones and cannot be filtered out as in the CW or mode-locked cases.

The correct estimation of the number of coincidences caused by the uncorrelated photons is crucial to measure entanglement in any setup using biphotons. In the best of our knowledge, the problems of the accidental coincidences in the well separated pulse pump regime, and also during (or inside) the pump pulse, have not been considered before. In this paper, we focus on the practical issues related with the measurement of the entanglement produced by SPDC in these cases. The methods to calculate the number of accidental coincidences in the several possible regimes are reviewed or derived in the Section \ref{secII}. These methods are then applied to measure the entanglement of the field produced when a pair of crossed BBO-I crystals is pumped with a UV Q-switched laser (pulses with a duration in the nanosecond [ns] range with a repetition rate in the order of kHz). The experiments are described in the Section \ref{secIII}, including an original application example: the estimation of the time variation of the Concurrence inside the pump pulse. The results and conclusions are commented in the Section \ref{secIV}.

\section{CALCULATING COINCIDENCES} \label{secII}
\subsection{Correlated coincidences} \label{secII1}  
In general, photons observed in detectors A and B at time values $t_{A}$ and $t_{B}$ are defined coincident if $\mid t_{A} - t_{B}\mid \leq T_{w}$, where $T_{w}$ is the duration of the so-called time window. The value of $T_{w}$ is arbitrary. In most experiments, it is fixed and given by the speed of electronic gates. These gates determine if the photon detections are coincident, or not, as the experiment runs. In the case of interest here, all the values of $t_{A}$ and $t_{B}$ are saved (what is named ``time-stamping'' or ``time tag'') so that the value of $T_{w}$ can be varied at will after the experiment has ended \cite{LG2004,aguero2009}. If the pump is pulsed, the time value of detection of the pulse is also saved.

To describe pulsed SPDC, the pump field is written as a superposition of monochromatic waves, so that the output SPDC state is an integral over the pulse spectrum: $|\Psi_{pulse}\rangle = \int d^{3}k |\Psi [F(k)]\rangle$ where $|\Psi [F(k)]\rangle$ is the output state produced by a CW pump with wave vector $k$ and field amplitude $F(k)$ \cite{joobeur1994,grice1997,kim2002}.
In the case of ns pulse pumping, the bandwidth $\Delta\omega _{pump}$ is smaller than the bandwidth of the SPDC process in the crystals and also smaller than the filters' bandwidth, so that the probability of detecting one photon of the SPDC pair at time $t$ and the other one at $t'$ is:
\begin{equation}
P(t,t')\propto \exp[-\Delta \omega^{2}_{filters} (t-t')^{2}]\times \exp\left[\frac{-\Delta \omega^{2}_{pump}(t+t')^{2}}{2}\right]
\label{ec1}
\end{equation}
which means that the time correlation of the detections is defined by the resolution of the spectral filters, while the events can only happen at the times dictated by the pump \cite{pan2012}. This defines the natural time $T_{nat}$, during which true coincidences are able to appear. The duration and timing of $T_{nat}$ is coincident with the duration and timing of the pump pulse. Coincidences observed outside $T_{nat}$ cannot be caused by the SPDC pair, and hence they are not correlated.
 
Once $T_{nat}$ is defined, the probability of a correlated coincidence is given by the probability of observing a photon in one of the detectors if a photon has already been observed in the other detector. This probability is affected by several practical imperfections: $\eta$ due to the detector's efficiency, $\varphi$ due to the transmission of the spectral filters, and $\gamma$ due to the geometry of the alignment (each of these three numbers is $\leq$ 1). The total probability of observing a correlated photon (provided that the other one has been already observed in the other detector) is then $\gamma\times\varphi\times\eta$. This factor relates the rate of single counts observed at one detector with the rate of observed correlated coincidences.
 
\subsection{Accidental coincidences, CW case} \label{secII2}
In the practice of the optical experiments with entangled states, a large number of uncorrelated photons are detected in addition to the correlated ones. Statistically, those produce accidental (spurious) coincidences that must be subtracted from the total number of recorded coincidences in order to evaluate the degree of entanglement achieved. A correct estimation of the number of accidental coincidences is hence crucial in any experiment of this type. In general, the probability of observing an accidental coincidence is the product of the single probabilities $P_{A}$ and $P_{B}$ of observing an event in each of the detectors. The total number of accidental coincidences in a given experimental run is then obtained after multiplying by the number of observations during that run: 
\begin{equation}
N_{acc}=P_{A}\times P_{B}\times N_{obs}.
\label{ec2}
\end{equation} 
In the case of CW pump, $P_{A} = N_{A}/N_{obs}$ (where $N_{A}$ is the number of single counts at detector A), the same for B, and $N_{obs}$ is the total time $T_{total}$ of the experimental run divided by the duration of the chosen time window $T_{w}$. The rate of accidental counts (i.e., the number of counts divided by the total time) for the CW case is then: 
\begin{equation}
R_{acc}=R_{A}\times R_{B}\times T_{w}.
\label{ec3}
\end{equation}  
Single photon detectors produce a nearly constant (in time) rate of counts even if they are not illuminated. This rate of dark counts $R_{d}$ is specified for each detector, and it typically ranges from 100 to 500 s$^{-1}$ for avalanche photodiodes cooled by Peltier cells (these are the detectors of most widespread use nowadays). The Eq. (\ref{ec3}) is always valid to calculate the accidental coincidences due to the dark counts. 
 
Minimizing the rate of accidental coincidences is obviously desirable and, in particular, it is unavoidable in the practice to allow the alignment of the setup (see also Section \ref{secII4}). In CW operation, the usual method to filter accidental coincidences out is to make $T_{w}$ as small as possible. In the pulsed operation instead, this method is limited, because most of the uncorrelated photons are emitted during the pump pulse, and hence simultaneous with the correlated ones. In order to calculate the rate of accidental counts in the pulsed regime, it is convenient considering two different cases: 
\\\textit{Case 1}: One wants to know (or is able to measure) the number of coincidences produced during the whole pump pulse, or $T_{w} \geq T_{pulse}$. This is the most usual case, and it includes the experiments using fs laser pumping, as entanglement swapping, teleportation, delayed-choice welcher weg, etc.  
\\\textit{Case 2}: When $T_{w}<T_{pulse}$. This is, in general, the case of relatively long ($>10$ ns) pump pulses. A particular and important sub-case is when one wants to know the temporal distribution of the coincidences within the pump pulse. This corresponds, f.ex., to the time-resolved measurement of the violation of the Bell's inequalities \cite{aguero2012} and, in general, to any time-resolved study involving photon coincidences. 
 
\subsection{Accidental coincidences: Case 1 ($T_{w} \geq T_{pulse}$)}\label{secII3}
 
In this case, $N_{obs}$ is the number of pulses during the run, or $T_{total} \times f_{rep}$ (where $f_{rep}$ is the laser's repetition rate). Therefore \cite{kuzucu2008}:
\begin{equation}
R_{acc}=R_{A}\times R_{B}/f_{rep} + R_{d}^{2}\times T_{w},
\label{ec4}
\end{equation}  
where the second term is the contribution due to the dark counts, which is usually small. It is assumed here that $R_{d}$ is equal for both detectors, otherwise, $R_{d}^{2}$ must be replaced by the product $R_{d}^{A}\times R_{d}^{B}$.  Note that $R_{A}$, $R_{B}$ are the single count rates at each detector observed inside $T_{nat}$ only (i.e., synchronous with and during the pump pulse). This is the expression valid when pumping with mode-locked lasers (as, f.ex., in entanglement swapping, teleportation, etc.), since the pulse duration (ps or fs) is much shorter than the resolution time of the detectors and electronics. Eq. (\ref{ec4}) is also valid when pumping with a Q-switched laser pulses (ns) if one is not interested in the evolution of the number of coincidences during or inside the pulse, but only in the total number of coincidences produced by the whole pulse (as, f.ex., in quantum key distribution in daylight).
 
\subsection{Accidental coincidences: Case 2 ($T_{w} < T_{pulse}$)}\label{secII4}
 
Consider the simplest case of an ideally square-shaped pump pulse and time coincidence window. The probability of observing a photon in detector A during $T_{w}$ is:
\begin{equation}
P(A)|_{T_{w}} = \frac{N_{A}}{N_{obs}}\frac{T_{w}}{T_{pulse}}=\frac{R_{A}}{f_{rep}}\frac{T_{w}}{T_{pulse}},
\label{ec5}
\end{equation}
where, as in the Case 1, $N_{A}$ is restricted to the number of single counts obtained during $T_{nat}$. The total rate of accidental counts (i.e., including dark counts) is therefore: 
\begin{equation}
R_{acc}=(R_{A}\times R_{B}/f_{rep})(T_{w}/T_{pulse})+ R_{d}^{2}\times T_{w}.
\label{ec6}
\end{equation}  
In the limit $f_{rep}\times T_{pulse}\rightarrow 1$, the Eq. (\ref{ec3}) for the case of CW pumping is retrieved. The Eq. (\ref{ec6}) allows defining the appropriate pump power level. At the end of the Section \ref{secII1}, it was stated that the rate of valid coincidences is: 
\begin{equation}
R_{valid}=R_{A}\times\gamma\times\varphi\times\eta,
\label{ec7}
\end{equation}   
where $\lbrace\gamma ,\varphi ,\eta\rbrace$ are the efficiency values for the B station. It was also stated that a high ratio $r$ between the valid and the accidental coincidences is necessary to align the setup. Neglecting the last term in Eq. (\ref{ec6}), which is usually small, and assuming that the two stations have similar efficiencies, the valid-to-accidental ratio can be then estimated as: 
\begin{equation}
r \approx (T_{pulse}/T_{w})\times\gamma\times\varphi\times\eta/p,
\label{ec8}
\end{equation} 
where $p\equiv R_{B}/ f_{rep}$ is the probability \textit{per pulse} of detecting a (single) photon in the detector B. Typical numbers are $\gamma\times\varphi\times\eta\approx 0.05$, and $T_{pulse} /T_{w}\approx 10$, so that $p\leq 0.05$ to get a value of $r$ that allows alignment.  
As a consequence, in this regime the pump intensity must be adjusted low enough so that most pulses do not produce a detected photon. Fulfilling this criterion has an additional advantage: one of the phenomena that spoil entanglement is the emission of two pairs of entangled photons during the same time coincidence window. The probability of such event is $\propto p^{2}$ which, in these conditions, amounts to a negligible contribution. 
\begin{figure}[b]
\includegraphics[scale=0.5]{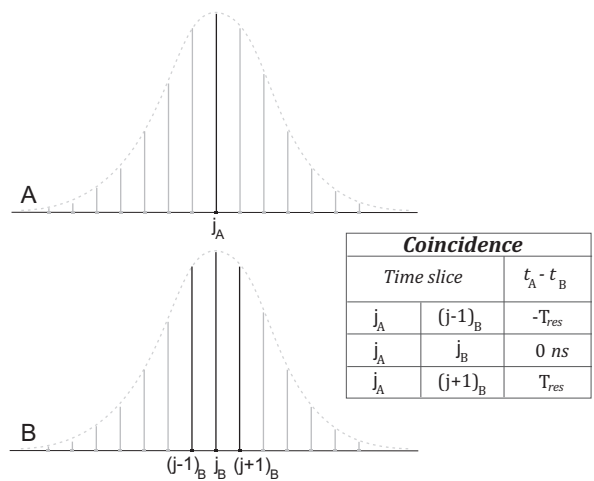}
\caption{\label{fig:ma1} Coincidence timing analysis. A detected photon at the slice $j_{A}$ of duration $T_{res}$ can produce a measured coincidence if a photon is detected at slices $(j-1)_{B}$ , $j_{B}$ or $(j+1)_{B}$. }
\end{figure} 

Measuring the time evolution of the coincidences inside the pump pulse is a special sub-case of the Case 2. The pulse is sliced in as many time intervals of duration $T_{w}$ as possible. The number of slices is then limited by the system's resolution $T_{res}$.  As the rate of single counts varies during the pulse, $R_{acc}$ must be calculated for each time slice (Fig. \ref{fig:ma1}). 
Note now that a detection at slice $j_{A}$ (this means: the j-slice of the set of data obtained at the A station) is at a distance $T_{res}$ from a detection at slice $(j-1)_{B}$, and also from a detection at slice $(j+1)_{B}$. Therefore, the correct rate of accidental coincidences in slice j inside the pulse is (neglecting for the moment the small contribution from the dark counts): 
\begin{equation}
R_{acc}(j)= \frac{R_{A}(j)\times [R_{B}(j-1)+R_{B}(j)+R_{B}(j+1)]}{f_{rep}}, 
\label{ec9}
\end{equation}  
i.e., in addition to the rate of detections at the same j-slice, the contributions from the previous and following slices must be taken into account.   
 
\section{MEASURING COINCIDENCES} \label{secIII}
In this section, we describe the application of the expressions obtained in the previous pages to an experiment producing biphotons using ns-pump pulses at kHz repetition rate.  

\subsection{The experimental setup} 

\begin{figure}[b]
\includegraphics[scale=0.38]{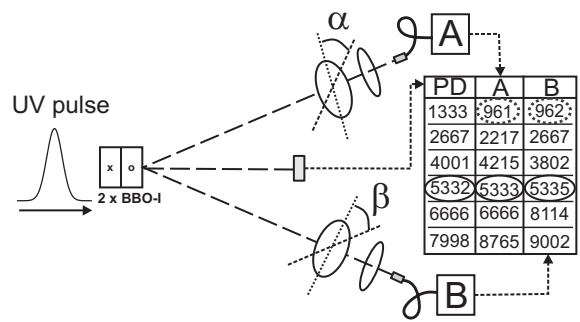}
\caption{\label{fig:ma2} Sketch of the experimental setup. The time of detection of each photon (A, B) and of each pulse (PD) are measured and saved with a resolution of 12.5 ns. The natural time for photon detection $T_{nat}$ is an interval of 75 ns (the full pulse duration) synchronous with the PD ``alert'' signal. All the delays and time coincidence windows can be varied at will after the experiment has ended. In the first line there is a coincidence ($T_{w}^{AB} \geq 25$ ns) between A and B, but it is outside $T_{nat}$ (assuming that all the delays are zero). A coincidence inside $T_{nat}$ is in the fourth line. }
\end{figure}

A set of two crossed BBO crystals cut for Type-I SPDC \cite{kwiat1995} is pumped with 355 nm radiation of the third harmonic of an actively Q-switched diode-pumped Nd:YVO$_{4}$ laser designed and built in this lab \cite{aguero2010}. At a repetition rate of 60 kHz, the pulses are 35 ns FWHM with a coherence length of 14 mm, much longer than the crystals (1 mm long each). The fully symmetrical Bell's state $|\phi^{-}\rangle$ is thus produced. The SPDC radiation is detected with silicon avalanche photodiodes operated in the Geiger mode. They are coupled through multimode optical fibers with a core of 100 $\mu$m and N.A.= 0.29. Before each module, there is an arrangement of optical elements composed of a variable diameter iris, followed by a polarization analyzer, an interference filter ($\Delta\lambda$ =10 nm at 710 nm), and a microscope objective focusing the 710 nm radiation into the multimode optical fiber (see Fig. \ref{fig:ma2}).

The time values of each single photon detection, as well as the ``alert'' signal provided by a fast photodiode detecting the pump pulse, are recorded with the help of a NI-6602 PCI board \cite{aguero2009}. The time-stamped series are saved in the hard disk of a PC, so that the values of delays and time windows (which determine $T_{nat}$ and hence the number of coincidences) can be varied at will after the experiment has ended. Therefore, output of an experimental run consists of three series of time stamped values: two for the photons detected at each station (A and B) and one for the pump pulse (PD). Be aware that, as the pump power is adjusted such that $p\ll 1$, the number of lines in the PD column is much larger than in the A or B ones. The NI-6602 board has a time resolution of 12.5 ns and three Direct Memory Access 32-bit counters, so that it is possible to record up to $\approx$ 53.68 s of real time information without interruption. We name ``ramp'' to each set of (three series of) data obtained in these conditions. One ramp is the building block of the data in these experiments. 

\subsection{Measuring accidental coincidences} 

The Eq. (\ref{ec6}) is derived for perfect square shaped pulses and coincidence windows. For real shapes, the factor $T_{w}/T_{pulse}$ must be replaced by a ``form factor'' given by a convolution between the pulse and time window shapes. In the practice, measuring the form factor is more precise and simpler than calculating it. This is done by measuring the number of accidental counts when purposely spoiling the correlation. In other words, Eq. (\ref{ec6}) is written now as:
\begin{equation}
R_{acc}=(R_{A}\times R_{B}/f_{rep})\times k_{0}+ b_{0},
\label{ec10}
\end{equation}  
where $k_{0}$ and $b_{0}$ are parameters that depend on the values of $T_{w}$ and $T_{pulse}$, and that are determined from auxiliary measurements. 

The obvious way to obtain uncorrelated coincidences is to misalign the setup. Yet, this procedure has the disadvantage of the uncertainty in the changes that are introduced in the factor $\gamma$ and the difficulty in recovering the correct alignment. A safer procedure is to place a piece of white paper just after the BBO crystals in the Fig. \ref{fig:ma2}. The UV pump radiation induces the chemicals in the paper to emit (synchronous with the pump pulse) uncorrelated fluorescence with a broad spectrum. 

\begin{figure}[b]
\includegraphics[scale=0.31]{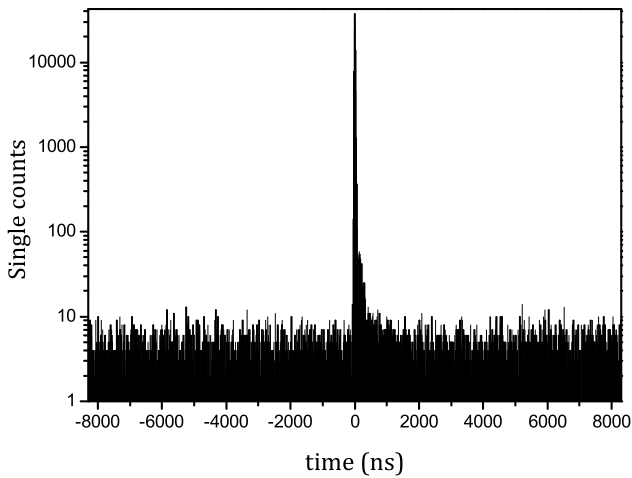}
\caption{\label{fig:ma3} Number of single counts in detector B as a function of the delay with the PD time list, $T_{w}=T_{res}=12.5$ ns.  }
\end{figure}

The next task is to find the value and position of the natural time of detection $T_{nat}$. The delay between the three time stamped lists (which may be different from zero because of different cable lengths, time response of the electronics, etc.) are adjusted from the plot of the number of single counts as a function to the distance to the peak of the pump pulse, which is defined as $t=0$ (Fig. \ref{fig:ma3}). The time position and width of the peaks in these figures (one for each station) define $T_{nat}$.

The inserted piece of paper also blocks the faint correlated SPDC radiation produced in the crystals. Therefore, in these conditions all the photons reaching the detectors are uncorrelated. This is further checked by measuring the coincidence rate (with the delays optimized as explained above) at different orientations $\lbrace\alpha,\beta\rbrace$ of the analyzers. As it is expected, no variation is observed. 
\begin{figure}[h!]
\includegraphics[scale=0.31]{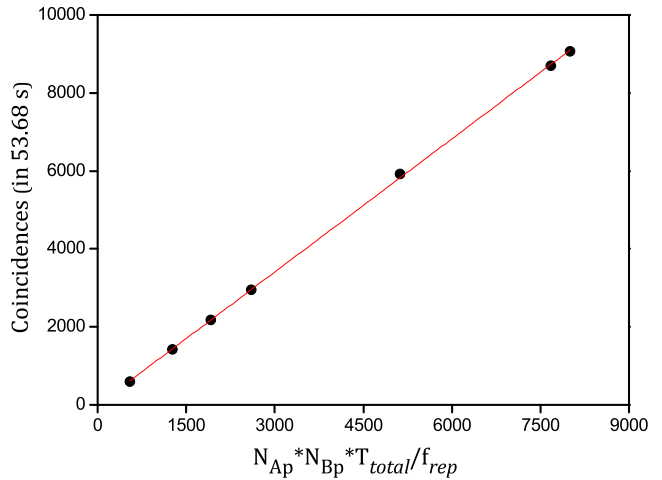}
\caption{\label{fig:ma4} Number of coincidences (after delay optimization) as a function of the product of the single counts  for $T_{w}=75$ ns. The linear fit provides the values $k_{0}= 1.139 \pm 0.007$ and $b_{0}=-0.1\pm 0.6$.  }
\end{figure}

In order to measure the parameters $k_{0}$ and $b_{0}$, the rate of single counts is changed by varying the apertures of the irises in front of the detector modules. The number of coincidences between the time stamped lists A and B containing events only inside $T_{nat}$ is recorded. To be precise: two detections A and B are considered coincident if $|t_{A}-t_{B}|\leq T_{w}^{AB}$ (after $t_{A}$ and $t_{B}$ have been corrected with the optimized delay values), A and PD if $|t_{A}-t_{PD}|\leq T_{w}^{APD}$ and B and PD if $|t_{B}-t_{PD}|\leq T_{w}^{BPD}$. In the Fig.  \ref{fig:ma4}, the results are displayed for the choosing $T_{w}^{APD} = T_{w}^{BPD} = T_{w}^{AB} = T_{nat} = 75$ ns. The number of coincidences, plotted as a function of the product of the single counts, allows (as expected) a linear fit. The values of $k_{0}$ and $b_{0}$ are found in this way. The procedure is repeated for $T_{w}$ values ranging from $T_{nat}$ (75 ns) down to the time resolution of the data acquisition system (12.5 ns). 
The obtained values for $k_{0}$ for each value of $T_{w}$ are summarized in the Table \ref{tab:table1}. In all cases $b_{0}$ (which is the contribution from the dark counts rate inside $T_{nat}$) is practically zero, as expected. This Table allows the calculation of the number of accidental counts as a function of the single counts in each station, for different values of the time coincidence window. Of course, these values are specific of our detectors and data acquisition system. In a different setup, the procedure should be repeated. 
\begin{table}[h!]
\caption{\label{tab:table1}%
Measured values of the ``form factor'' $k_{0}$ for different values of the time coincidence window.}
\begin{tabular}{cc}
\hline \hline
\textrm{$T_{w}$(ns)}\hspace*{1cm}&
\textrm{$k_{0}$}\\
\colrule
12.5 \hspace*{1cm}& 0.658 $\pm$ 0.002 \\
25 \hspace*{1cm}& 0.938 $\pm$ 0.003 \\
37.5 \hspace*{1cm}& 1.069 $\pm$ 0.003 \\
50 \hspace*{1cm}& 1.112 $\pm$ 0.003  \\
62.5 \hspace*{1cm}& 1.126 $\pm$ 0.003  \\
75  \hspace*{1cm}& 1.139 $\pm$ 0.007 \\ \hline \hline
\end{tabular}
\end{table}

F.ex.: the total number of single counts in the Fig. \ref{fig:ma3} (one ramp, $T_{total}$= 53.68 s) is 45072 (file: S9D11552). After optimizing the delay, the number inside Tnat (i.e., inside the central peak 75 ns wide) is 33745. The total single counts in detector B (file: S9D21552) are 127377, and 120425 inside $T_{nat}$. Then, from Eq. (\ref{ec10}) with the value of $k_{0}$ from Table \ref{tab:table1} for $T_{w}=75$ ns, the number of accidental coincidences inside $T_{nat}$ is 1437. The directly measured number (i.e., coincidences in the three time stamped files) is 1417, which is coincident with the calculated one within the statistical fluctuation.  By the way, the number of recorded pump pulses in this ramp (file: S9D31552) is 3234432. The complete set of raw experimental data is available in the website \cite{pagweb}.

 In order to know the time evolution of the correlation inside the pump pulse, it is necessary to estimate the number of accidental coincidences in each time slice. According to Eq. (\ref{ec9}), this requires knowing the rate of single counts $R_{A}(j)$ and $R_{B}(j)$ in each j-slice. Therefore, we calculate the coincidences between lists A, B and PD ($T_{w}^{APD} = T_{w}^{BPD} = T_{nat} = 75$ ns  and $T_{w}^{AB} = T_{res} = 12.5$ ns) and also record the position, with respect to the peak of the pump pulse, of the time slice where each coincidence occurs. F.ex., for the time slice in the center of the pulse and after optimizing the delays, the number of single counts is 10479 (detector A) and 37247 (detector B). The previous and following slices have 5990 and 8769 counts (detector A) and 24379 and 29334 (detector B) respectively. From Eq. (\ref{ec9}), the number of accidental coincidences for the time slice in the center is 293 (note: this is the average value between $R_{A}(j)\times [R_{B}(j-1)+\cdots]$ and $R_{B}(j)\times[R_{A}(j-1)+\cdots]$; anyway, in all the files we studied   the difference between these two results is found negligible). The actual number of coincidences measured at that slice is larger: 336. The procedure is repeated for all the time slices, and a histogram of the number of accidental coincidences inside the pump pulse is calculated.  
\begin{figure}[h!]
\includegraphics[scale=0.31]{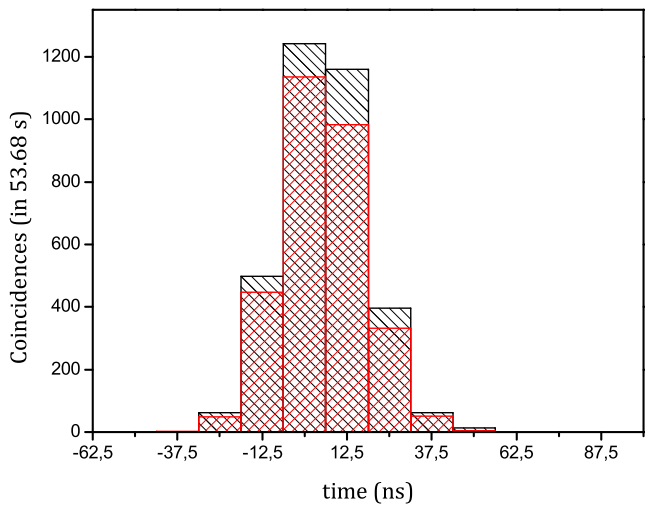}
\caption{\label{fig:ma5} Distribution of accidental coincidences inside the pump pulse: $\backslash\backslash\backslash$ (black) directly measured with the piece of paper inserted, /// (red) calculated from the number of single counts and Eq. (\ref{ec9}).}
\end{figure}

In the Fig. \ref{fig:ma5}, the time distribution of the accidental coincidences calculated from Eq. (\ref{ec9}) by using the numbers of single counts is plotted (///) (red online). It is also plotted ($\backslash\backslash\backslash$) the number of coincidences in each time slice as directly measured by counting triple coincidences in the recorded data files. 
The values calculated from the Eq. (\ref{ec9}) are, for all the values of time, smaller than the directly measured ones. The deviation is larger at the center of the pulse, and negligible at the edges. The conclusion is that the Eq. (\ref{ec9}) underestimates the actual number of accidental coincidences, about a 10$\%$ at the center of the pulse. The consequence is a systematic underestimation of the real entanglement. In spite of this drawback, the Eq. (\ref{ec9}) provides a simple and satisfactorily accurate tool to calculate the number of accidental coincidences during the pulse, and it is used in what follows.

\begin{figure*}[htp]
\subfigure[]{\includegraphics[scale=0.28]{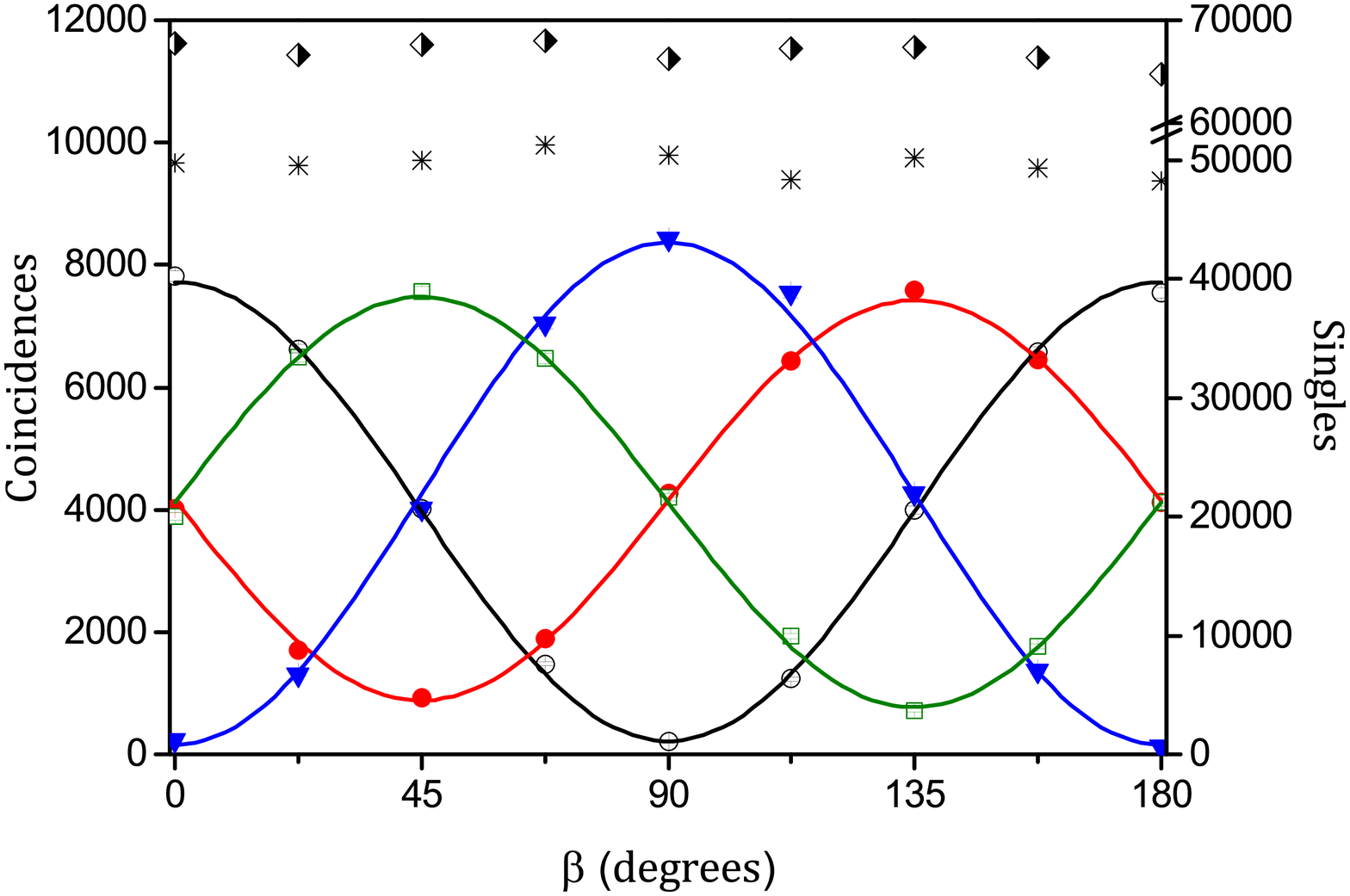}}
\subfigure[]{\includegraphics[scale=0.25]{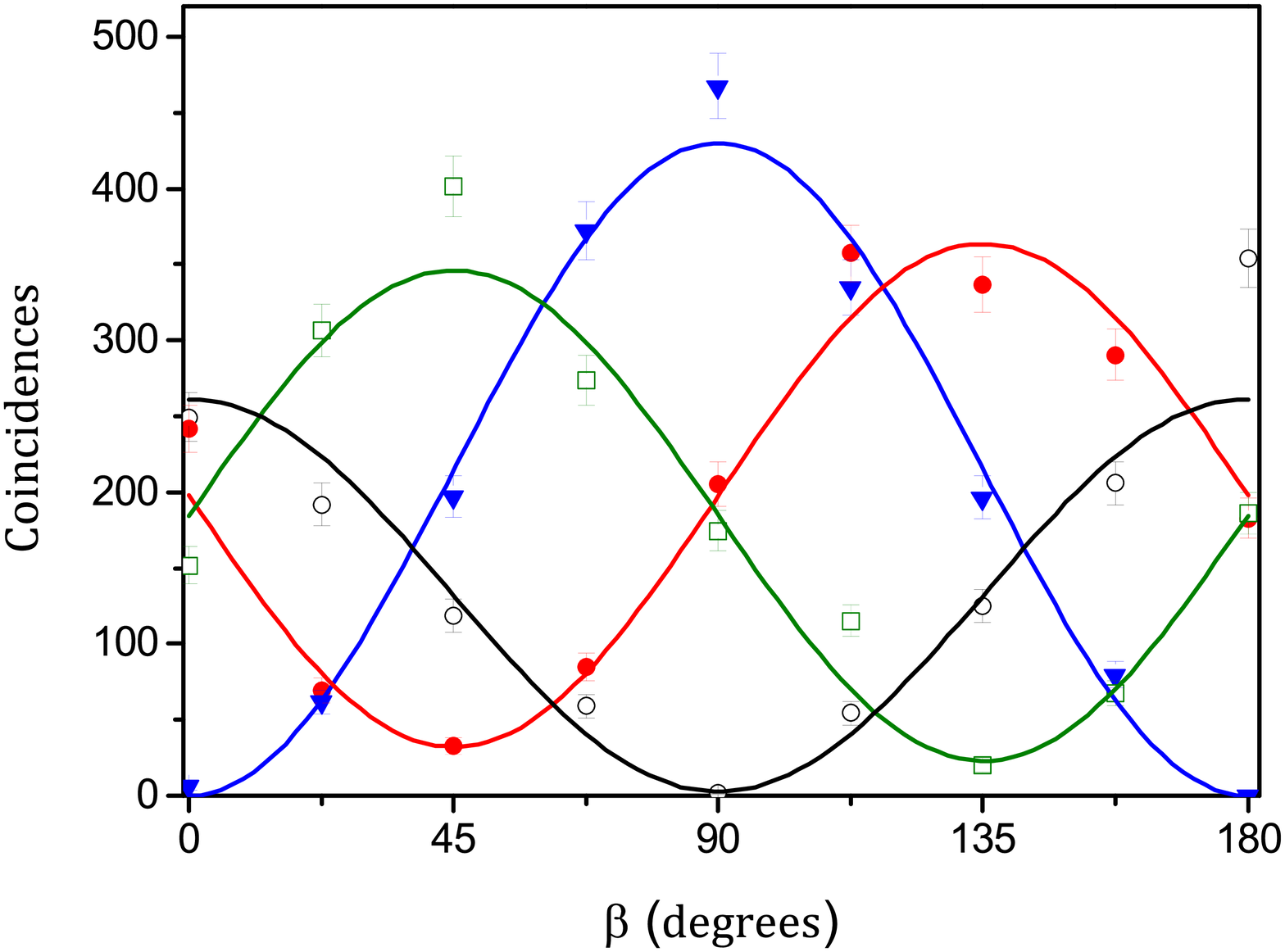} }
\subfigure[]{\includegraphics[scale=0.25]{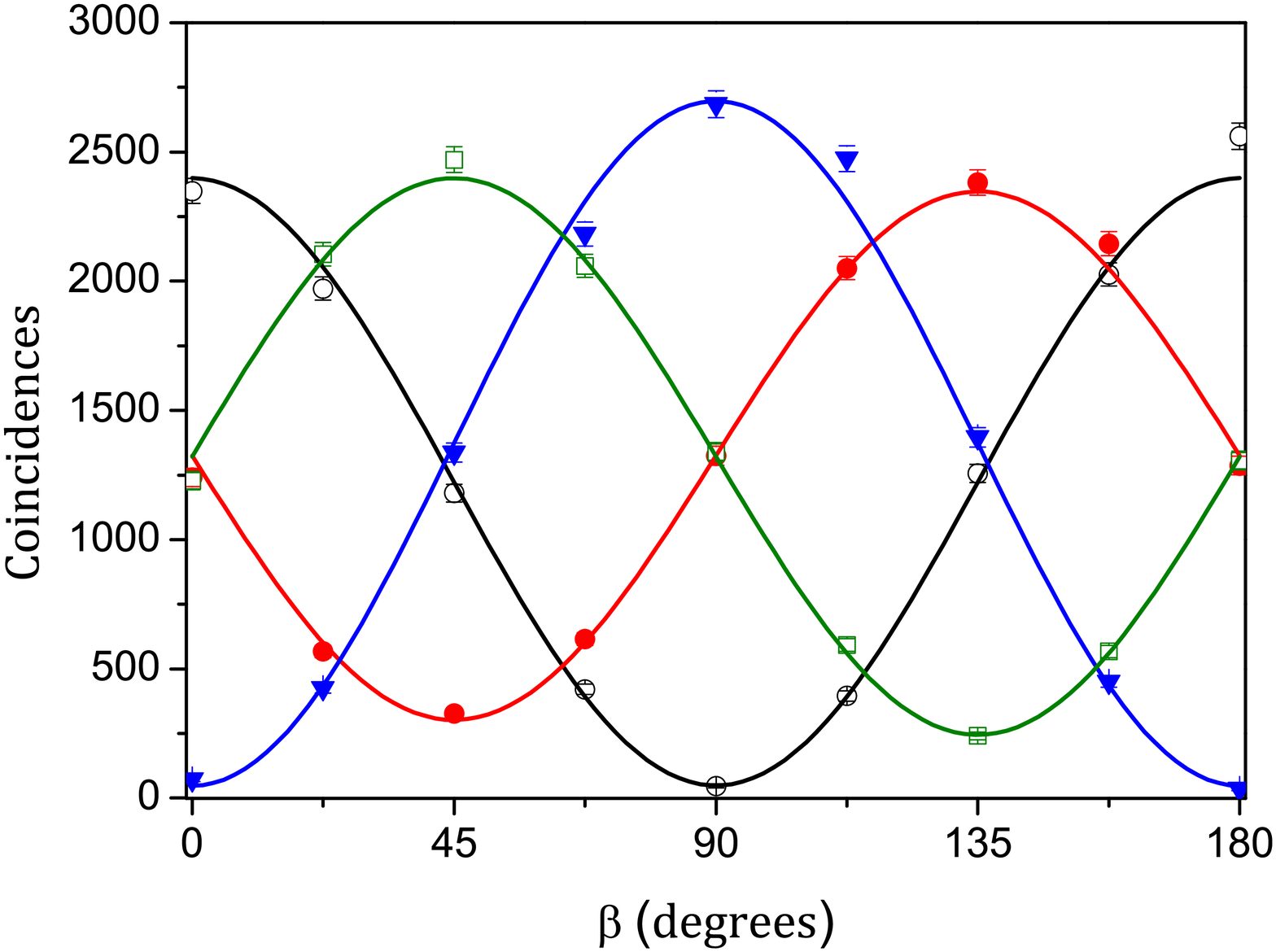}} 
\hspace{0.6cm}
\subfigure[]{\includegraphics[scale=0.25]{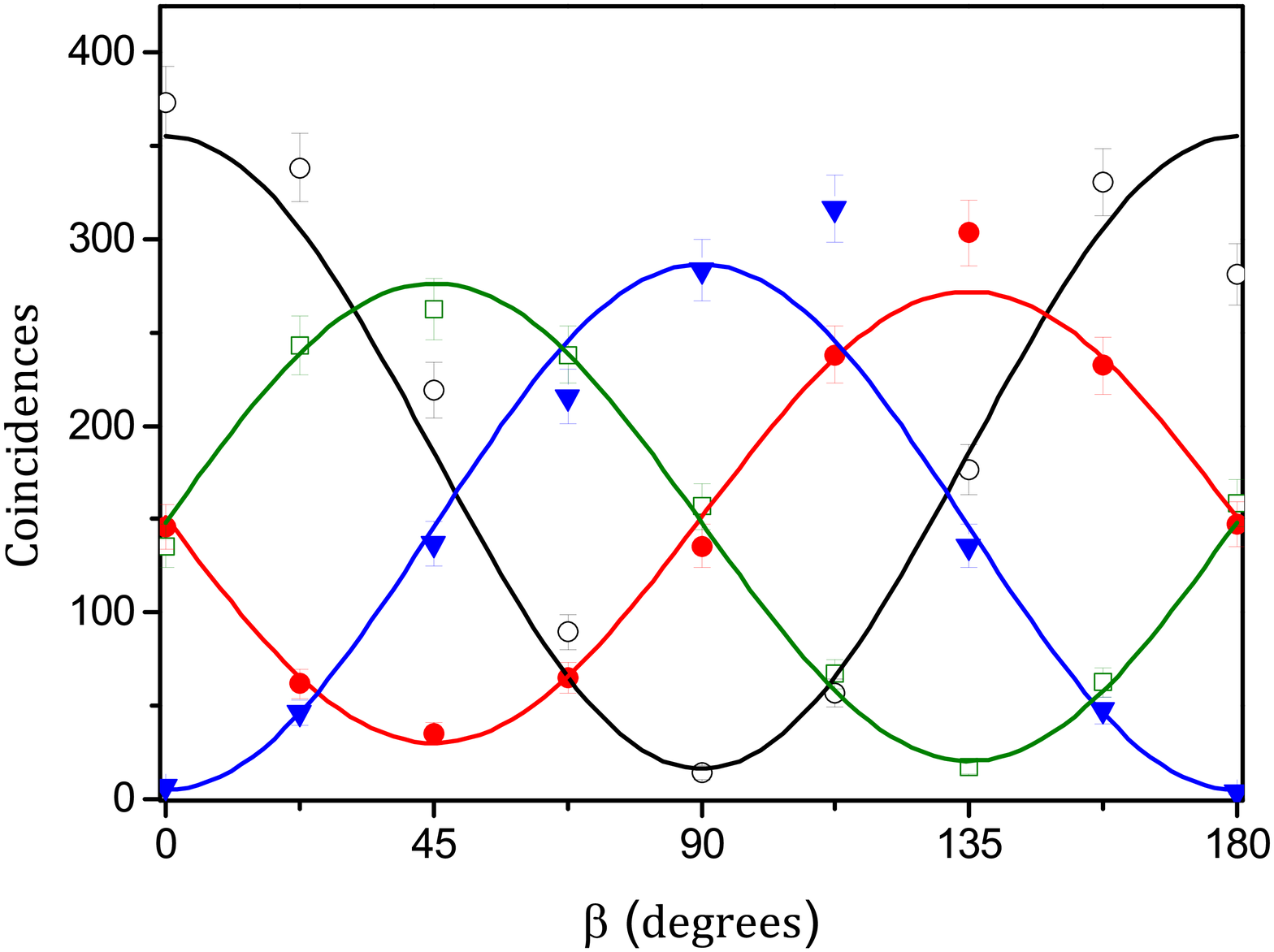}}
\caption{\label{fig:ma6} Number of correlated coincidences (scale on the left) as a function of the analyzers' angles: open circles (black) $\alpha=0^{\circ}$, full circles (red) $\alpha=45^{\circ}$, triangles (blue) $\alpha=90^{\circ}$, squares (green) $\alpha=135^{\circ}$; (a): for the full pulse duration, rhombs (detector A) and asterisks (detector B) indicate the number of single detections (scale on the right); (b-d) for three time slices: $t=-25$ ns (b), $t=0$ (pulse peak) (c), and $t=37.5$ ns (d). }
\end{figure*} 

The piece of paper is now removed, and time-stamped files of photon detections are recorded for different orientations $\lbrace\alpha,\beta\rbrace$ of the analyzers. The delays are optimized, and the number of accidental coincidences is calculated and subtracted from the total number of measured coincidences. The result is the number of valid, or correlated, coincidences. By this procedure, it is possible to draw curves of valid coincidences for the whole pulse ($T_{w}=T_{pulse}$, Fig. \ref{fig:ma6}a) and also for each time slice of 12.5 ns inside the pump pulse (a few examples in Figs. \ref{fig:ma6}b-d).
 
F.ex., for $\alpha=45^{\circ}$, $\beta=135^{\circ}$ (files: S7D11640, S7D21640 and S7D31640) and the time slice in the center of the pulse, the number of singles are 22010 for detector A and 16859 for detector B, with the previous and next slices values 11350 and 12387. From Eq. (\ref{ec9}) the number of accidental counts is estimated 277, which are subtracted from the total coincidences at that slice (2659), providing 2381 of valid counts, plotted as the dot at the maximum of the full circles (or red) curve in the Fig. \ref{fig:ma6}c. This means a value of $\gamma\times\varphi\times\eta\approx$ 0.14, which is measured nearly constant during the pulse. In the Fig. \ref{fig:ma6}a the total (in time) numbers of single counts in each detector are also plotted (scale on the right) to further show that there is no dependence with the analyzers' orientation. Note the different vertical scales, due to the different instantaneous pump intensity at the different time slices. Note also that the fitting to the theoretical curves improves as the number of coincidences increases. 

\subsection{An application example: measuring the time variation of the Concurrence}

In \cite{aguero2012}, the procedure described in the previous pages was applied to obtain the time-resolved evolution of the Clauser-Horne-Shimony and Holt parameter ($S_{CHSH}$) inside the pump pulse. This parameter indicates whether one of the Bell's inequalities is violated or not, but it does not provide, strictly speaking, a measure of entanglement. As an illustrative example of the application and possibilities of the described techniques, in this Section we calculate the time evolution of the Concurrence. This example adds nontrivial information to the results in \cite{aguero2012}, for it uses a different theoretical approach and also a different set of experimental data. In \cite{aguero2012}, the set of data consisted of 7 ramps for each of the 16 values of $\lbrace\alpha,\beta\rbrace$ necessary to calculate $S_{CHSH}$. Here, instead, the set of data consists of one ramp for each of 36 equally spaced angle values. 

One of the intriguing attributes of entanglement is that, in some conditions, it can show a discontinuous evolution. The abrupt vanishing of entanglement under the action of a decoherent channel has been named ``entanglement sudden death'' \cite{yu2006}. It has been observed by measuring the Concurrence in an optical setup equivalent to an environment with adjustable dissipation \cite{almeida2007}. In general, measuring Concurrence requires determining the density matrix through the procedure known as state tomography \cite{james2001}. This involves long series of correlation measurements and rather complex numerical methods, to make the experimental data (which are affected by statistical variations, drifts and noise) compatible with a physically meaningful density matrix. Nonetheless, if the form of the state is \textit{assumed}, the Concurrence is much simpler to measure. This simplified approach is taken here with the aim to get a quick glimpse on the general features of its time behavior, what suffices for the purposes of this application example. Let hence assume that the state of the field at the detectors in the Fig. \ref{fig:ma2} has the form of the mixture:
\begin{equation}
\rho=\kappa |\phi^{-}\rangle\langle\phi^{-}|+\mu \textbf{I} + \xi |\phi^{+}\rangle\langle\phi^{+}|,
\label{ec11}
\end{equation} 
where the first term is the contribution due to the aimed biphoton state, the second one is the contribution due to uncorrelated photons (\textbf{I} is the identity matrix), and the third one is the contribution due to the residual distinguishability or phase mismatch that may exist between the amplitudes produced in each of the two crystals. It is reasonable to expect that these three contributions are the main components of the actual state produced in the setup in Fig. \ref{fig:ma2}. Under this assumption (see the Appendix):
\begin{equation}
Concurrence = max(0,2\kappa +\mu /2-1).
\label{ec12}
\end{equation} 
The coincidence probability corresponding to the state in Eq. (\ref{ec11}) is:
\begin{equation}
P^{++}(\alpha,\beta)= \tfrac{1}{2}\kappa\thinspace cos^{2}(\alpha+\beta) +\tfrac{1}{4}\mu + \tfrac{1}{2}\xi\thinspace cos^{2}(\alpha-\beta).
\label{ec13}
\end{equation} 

\begin{figure}[h!]
\includegraphics[scale=0.28]{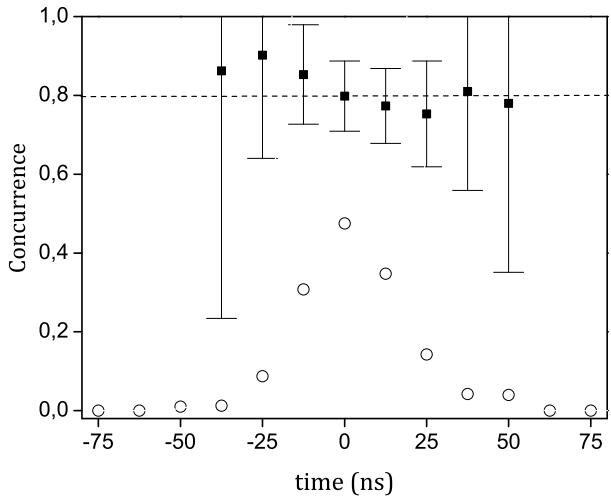}
\caption{\label{fig:ma7} Black squares: Concurrence (with error bars) as a function of time. Open circles: single detections in A (arbitrary vertical scale) to indicate the position and shape of the pump pulse. The horizontal dashed line indicates the value calculated using all the data inside the pulse (or Case 1).}
\end{figure}

The values of $\lbrace\kappa,\mu,\xi\rbrace$ are then obtained by the best numerical fit of $P^{++}(\alpha,\beta)$ to the set of data  recorded for the full pump pulse (Fig. \ref{fig:ma6}a), and also for each time slice (as the ones in Figs. \ref{fig:ma6}b-d). 
The curves in the Fig. \ref{fig:ma6} are drawn to guide the eye, the numerical fit of $P^{++}(\alpha,\beta)$ to measure $\lbrace\kappa,\mu,\xi\rbrace$ is done directly to the data. The value obtained for the full pulse (from Fig. \ref{fig:ma6}a, or Case 1) is $C= 0.795 \pm 0.051$, and it is indicated as a dashed horizontal line in the Fig. \ref{fig:ma7}. The values for the time variation of the Concurrence are displayed as black squares. The errors are calculated from the statistical fluctuation and standard error propagation. The pump pulse shape, which is obtained from the number of single counts in each time slice, is drawn as open circles to provide the time reference.

The Concurrence is measured practically constant during the time the pump is nonzero. The small ``bump'' downwards near the pulse's peak is probably caused by the slight underestimation of the accidental coincidences mentioned in the previous Section. The error bars naturally increase near the pulse's edges, where the statistics is scarcer. Yet, the expectation value remains constant within the error. The number of valid coincidences before -37.5 ns and after 50 ns is too small to allow a meaningful calculus, so that the Concurrence values and error bars are not plotted for those points. 
Anyway, at those extreme time slices (and beyond) the number of coincidences show no detectable modulation with the analyzers' angles, hence $\mu\approx 1$ $\Rightarrow$ $\kappa,\xi\ll 1$, and in consequence it is possible to state (from Eq. (\ref{ec12})) that the Concurrence for those slices is zero. Therefore, the measurable Concurrence falls to zero abruptly (within the available time resolution) at the pulse's end on the right, and it also appears immediately at the pulse's start on the left. 

We remark that not only the theoretical basis but also the experimental data used for this example are different from the ones used in the study reported in \cite{aguero2012}, so that entirely new information is presented here. Besides an illustration of the method of calculation of accidental coincidences, the Fig. \ref{fig:ma7} thus provides new experimentally based insight into the properties of entanglement.

\section{SUMMARY} \label{secIV}
We have presented the theoretical background and the practical methods to calculate the number of accidental coincidences for a pulsed source of biphotons. This number is essential to calculate the degree of entanglement achieved. Two different regimes are identified: \textit{Case 1}, when the pulses are shorter than the resolution time of the data acquisition system (or, if the observer is not interested in a time scale shorter than the pulse duration), and \textit{Case 2}, when the pulses are longer than the resolution time. This includes the special sub-case of measurements during or inside the pulse. 

The methods have been tested in an experimental setup producing biphotons through SPDC by pumping the nonlinear crystals with pulses of 75 ns total duration emitted by an actively Q-switched all-solid-state laser at a repetition of 60 kHz. The time of detection of each photon and of all the pump pulses (i.e., regardless if they produce detected photons or not) are recorded in separated time-stamped files with a resolution of 12.5 ns. This allows varying at will the delays between the series, and also the time coincidence windows, after the experiment has ended. The methods for the Case 1 and Case 2 are found accurate within the statistical fluctuation. The method to measure the time evolution inside the pulse underestimates the number of accidental coincidences, being the error largest ($\approx 10 \%$) at the peak of the pulses. The method thus leads to a slight underestimation of the entanglement, but it is anyway regarded as satisfactory.

 As an illustration of the described methods, we apply them to calculate the Concurrence for the full pulse duration and also for the time variation inside the pulse. The latter shows the expected discontinuous transition at the pulse's edges. This result is additional to, and independent from, the ones presented in \cite{aguero2012} for the time variation of the $S_{CHSH}$ parameter.

We expect that the methods described in this paper will find immediate application in the many experiments aimed to observe the time variation of any kind of correlation between photon detections (i.e., not only the correlation due to entanglement), where the practical estimation of the rate of accidental coincidences is necessary.

\section*{ACKNOWLEDGEMENTS}

This work received support from the contracts PIP08-2917 and PIP11-0077 of CONICET, Argentina.

\appendix

\section{Calculus of the Concurrence}
For the case of two qubits, the Concurrence can be calculated as \cite{wooters1998}:
\begin{equation}
Concurrence (\rho) \equiv C = max(0,\sqrt{\lambda_{1}}-\sqrt{\lambda_{2}}-\sqrt{\lambda_{3}}-\sqrt{\lambda_{4}}).
\label{ec13}
\end{equation} 
The $\sqrt{\lambda_{i}}$ are the square roots of the positive eigenvalues of the matrix \boldsymbol{$\rho.\rho'$} in decreasing order, where:
\begin{equation}
\boldsymbol{\rho' = (\sigma^{a}_{y}\otimes\sigma^{b}_{y})\rho^{*}(\sigma^{a}_{y}\otimes\sigma^{b}_{y})},
\label{ec14}
\end{equation} 
where \boldsymbol{$\sigma^{a(b)}_{y}$} is the spin-flip Pauli matrix acting in the a(b)-subspace. Due to the form of the state assumed in Eq. (\ref{ec11}), \boldsymbol{$\rho'=\rho$}. Using that $\kappa +\mu +\xi=1$, the eigenvalues of \boldsymbol{$\rho^{2}$} are:
\begin{equation}
\lambda_{i}=\lbrace(\mu /4)^{2}(twice),(\kappa+\mu /4)^{2},(\kappa -1 +3\mu /4)^{2}\rbrace
\label{ec15}
\end{equation} 
as $\kappa +3\mu /4 = 1-\xi -\mu /4 \leq 1$, then:
\begin{equation}
C=max(0,2\kappa +\mu /2 -1),
\label{ec16}
\end{equation}
which is the Eq. (\ref{ec12}). In general, $C\approx 1$ if $\kappa\approx 1$ ($\Rightarrow\mu ,\xi\approx 0$), and $C= 0$ if $\mu\approx 1$ ($\Rightarrow\kappa ,\xi\approx 0$). If $\xi = 0$, then $\mu=1-\kappa$ and $C= (3\kappa/2-1/2)$. Hence, even in this ideal condition, a minimum contrast $\kappa>1/3$ is necessary to get $C\neq 0$.

\bibliography{apsnacc}

\end{document}